\documentclass[10pt,twocolumn,amssymb, amsmath,nobibnotes,nofootinbib,aps,pre,showpacs,floatfix]{revtex4-2}       
\usepackage{graphicx}
\usepackage{multirow}
\usepackage{bm}
\usepackage{float}
\usepackage{dcolumn}
\usepackage[hidelinks]{hyperref}
\usepackage[dvipsnames]{xcolor}
\usepackage{silence}
\usepackage{booktabs}
\begin{document}
\title{Weak Itinerant Ferromagnetism (WIFM) in MAX phase compound Cr$_{1.9}$Fe$_{0.1}$GeC}
\author{Suman Mondal$^1$}
\email{mondals@post.bgu.ac.il}
\author{Mohamad Numan$^2$}
\author{Kurt Kummer$^3$}
\author{Sawada Masahiro$^4$}
\author{Subham Majumdar$^2$}
\affiliation{$^1$1Department of Physics, Ben-Gurion University of the Negev, Be’er-Sheva 84105, Israel}
\affiliation{$^2$School of Physical Science, Indian Association for the Cultivation of Science (IACS), 2A \& B Raja S. C. Mullick Road, Jadavpur, Kolkata 700 032, India}
\affiliation{$^3$ESRF European Synchrotron, 71 Avenue des Martyrs, F-38000 Grenoble, France}
\affiliation{$^4$Hiroshima Synchrotron Radiation Center (HiSOR), Hiroshima University-2-313 Kagamiyama, Higashi-Hiroshima 739-0046, Japan}

\begin{abstract}

Magnetic MAX phase compounds are important materials for studying the two-dimensional magnetism because of their layered crystallographic structure. The hexagonal MAX phase compound Cr$_2$GeC is a Pauli paramagnet, and here we report the induction of an ordered magnetic state by doping Fe at the Cr site. Induced magnetism for small doping concentrations (indicated as 5\% and 2.5\%) is found to have a weak itinerant ferromagnetic character. The Rhodes-Wolhfarth ratio is found to be 13.29, while the coefficient of electronic heat capacity ($\Gamma$) is 27 mJ-mol$^{-1}$K$^{-2}$ for Cr$_{1.9}$Fe$_{0.1}$GeC. Our x-ray magnetic circular dichorism measurement confirms that the magnetic moment arises from the Fe atom only, and Cr has negligible contribution towards the ordered moment. Our critical analysis indicates that the magnetic phase transition in Cr$_{1.9}$Fe$_{0.1}$GeC follows mean field theory.
\end{abstract}
\maketitle

\section{Introduction}

\begin{figure*}[htbp]
	\centering
	\includegraphics[width = 14 cm]{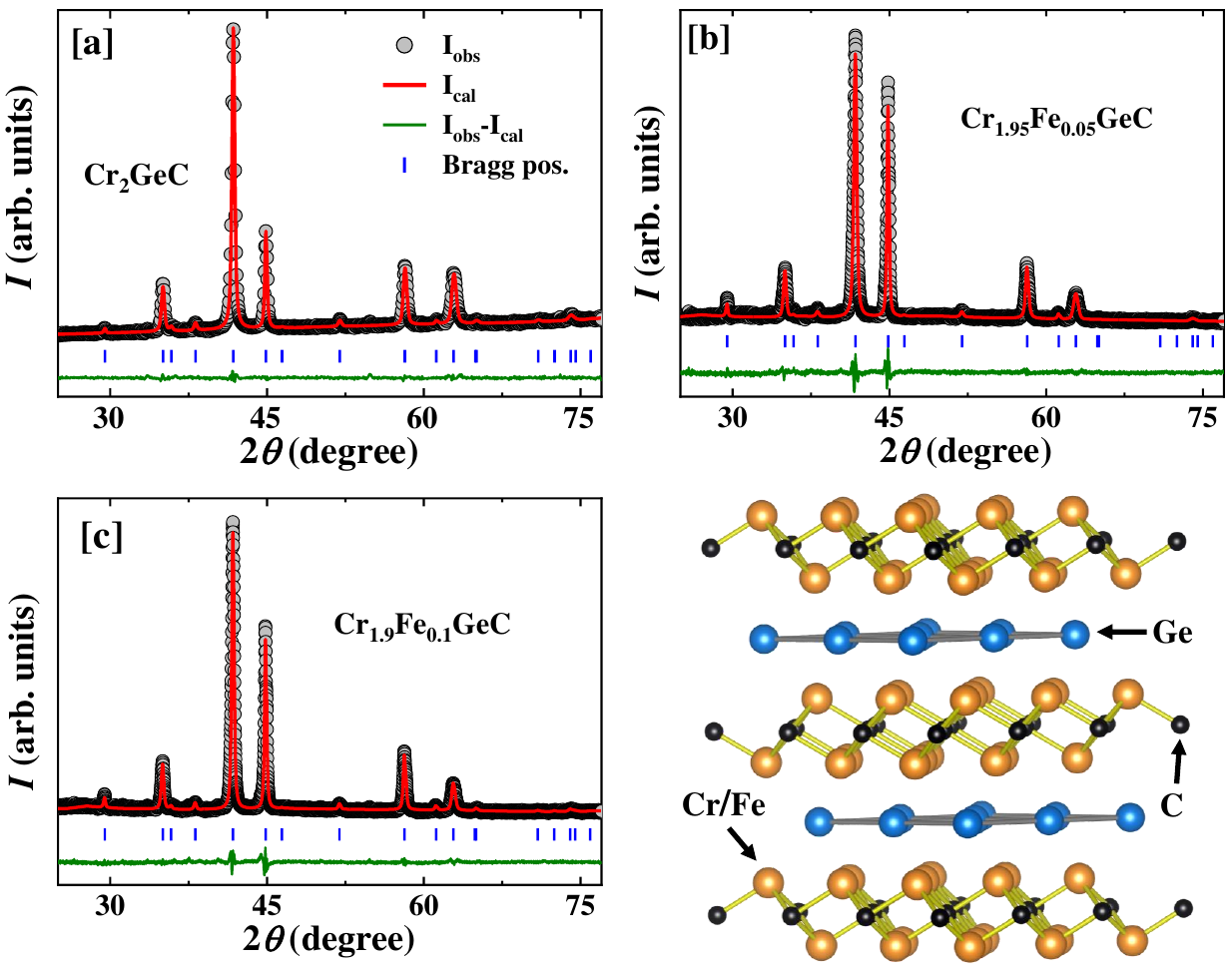}
	\caption {(a),(b) and (c) represent the PXRD patterns recorded at room temperature for $x =$ 0.0, 0.05 and 0.1 samples, respectively. The solid red lines through the data points depict the Rietveld refinement curves. (d) A perspective view of the crystal structure of Cr$_2$GeC-type MAX phase compound.}
	\label{fig:xrd}
\end{figure*}

Over the last few decades, there has been a growing activity in low-dimensional systems exhibiting layered sheet like structures such as graphene~\cite{han}, transition metal di-chalcogenides (TMDs)~\cite{nair}, layered van-der-Waals (vdW) materials~\cite{ahn}  due to their potential application in the next generation beyond silicon electronics. In particular, renewed interest in thermodynamically stable exfoliable nanolaminates, known as MAX phases with a general formula M$_{n+1}$AX$_n$ (n = 1, 2, 3) has been attributed to their intriguing physical, mechanical and chemical properties.  Furthermore, they also show a coexistence of both metallic and ceramic characteristics, reversible deformations, and high chemical resistance, among others~\cite{Barsoum,wang2009,Eklund,Sun,Radovic,Barsoum2003,Naguib}. Often MAX phases are characterized by a hexagonal layered structure, consisting of alternating layers of metal carbide or nitride (M$_6$X, M is an early transition metal and X = C or N) integrated with an $sp$-element (A) of the periodic table [see the crystal structure depicted in Fig.~\ref{fig:xrd}]. 
\par
Magnetic materials with layered structures are important from the fundamental as well as application point of view. The exchange coupling among the various layers is fascinating from the perspective of  spintronics and magnetic memory devices. Theoretically works indicate possible spin polarization is predicted in few MAX phase material, for example Cr$_2$AlC, Cr$_2$GeC etc ~\cite{Du,Ramzan,Dahlquist,Zhou,Mattesini}. However, the first ferromagnetic (FM) MAX phase Cr$_{2-x}$Mn$_x$AlC was successfully synthesized in the form of epitaxial thin films by Ingason \textit{et al.}~\cite{Ingason} in 2013. Subsequently, ferrimagnetic ordering was observed in the CrMnGaC thin film~\cite{Lin}. Furthermore, Mn-rich FM state and Mn-poor reentrant cluster glass state in  polycrystalline Cr$_{2-x}$Mn$_x$GeC samples~\cite{tao} were reported. Nakamura \textit{et al.} found the spin density wave state in the nitride MAX phase Cr$_2$GaN and the Pauli paramagnetic state in carbide MAX phase Cr$_2$GaC~\cite{Liu}. A comprehensive study utilizing powder neutron diffraction and x-ray magnetic circular dichroism (XMCD) experiments revealed a rich magnetic phase diagram and intricate magnetic structures in the rare-earth-based MAX phase compounds (Mo$_{2/3}$Dy$_{1/3}$)$_2$AlC and (Mo$_{2/3}$Ho$_{1/3}$)$_2$AlC~\cite{i-max}.

\par
The carbide MAX phase compounds mostly show good metallic character with resistivity as low as few $\mu\Omega$-cm at the liquid helium temperature. The induced moment on doping magnetic elements at the M-site is found to be low~\cite{Lin2016}, and they can be possible candidates for weak itinerant ferromagnet (WIFM)~\cite{santiago}. In a metallic magnetic system, WIFM is accompanied by low magnetic moment and generally obeys the Stoner-Wohlfarth criteria~\cite{blanco}. The delocalized nature of the $d$ orbital in transition-metal-based intermetallic alloys and compounds plays an important role in itinerant magnetism. WIFMs show various fascinating properties such as spin fluctuations, quantum criticality, and non-Fermi liquid behavior~\cite{saxena,Pffeiderer,Uhlarz}. Due to the delocalized nature of the moment and spin-fluctuation, the self-consistent renormalization (SCR) theory is found to be more appropriate to describe WIFMs~\cite{Moriya}. WIFMs lie very close to the magnetic non-magnetic phase boundary, and often a small perturbation can give rise to a large change in the electronic and magnetic properties.

\par
In the present work, we intend to induce magnetism by Fe doping in the MAX phase compound Cr$_2$GeC, which was previously reported to be a Pauli paramagnetic material. We doped Fe at the Cr site to prepare the solid solutions Cr$_{2-x}$Fe$_x$GeC ($x=$ 0.05 and 0.1), and they turned out to be WIFM, with ferromagnetic Curie  points lying below the room temperature.

\section{Crystal structure and methodology}
Polycrystalline samples of Cr$_{2-x}$Fe$_x$GeC ($x=$ 0.0, 0.05 and 0.1) were prepared by the standard solid-state reaction route. Stoichiometric amounts of Cr, Fe, Ge and C were mixed in intimately and sealed in an evacuated quartz tube (vacuum level~$10^{-5}$ mbar). Subsequently, it was fired at 1223 K for 1 day. The powders were then pressed into pellets and heated in an evacuated quartz tube at 1273 K with several intermediate grindings. The structure and phase purity of the samples were investigated by powder x-ray diffraction (PXRD) using Cu K$_\alpha$ radiation. All reflections could be indexed on the basis of the hexagonal symmetry (space group $P6_3$/mcm, no. 193) for all three samples. Rietveld refinement (solid red line: fitting curve) of the XRD data were performed using the MAUD software package~\cite{MAUD}, as depicted in Fig.~\ref{fig:xrd} (a), (b) and (c). The Wyckoff positions are given in Table~\ref{tab:lat}. The obtained lattice parameters for Cr$_2$GeC are $a=$ 2.955 \AA~ and $c=$  12.107 \AA~, which match well with the literature~\cite{tao}. With an increase in Fe concentration, there is a small but systematic increase in both $a$ and $c$.
\par
The dc magnetization ($M$) of the samples was measured using a Quantum Design SQUID magnetometer (MPMS3)  in the temperature ($T$) region of 2 to 380 K and in the magnetic field region 0 to 50 kOe. The resistivity ($\rho$) was measured using the four-probe technique in the temperature range of 5-300 K in a laboratory setup containing a closed-cycle helium refrigerator. The heat capacity of the $x=$ 0.1 sample was measured using the relaxation technique in the Physical Property Measurement System (PPMS, Quantum Design Inc., USA), in the temperature range of 2 to 300 K. 

\par
X-ray absorption spectroscopy (XAS), as well as X-ray magnetic circular dichroism (XMCD)  experiments at the Cr L$_{2,3}$ and Fe L$_{2,3}$ edges, were carried out at the D32 beamline, ESRF~\cite{brookes}. The experimental end station allows one to reach magnetic fields up to 90 kOe and temperature on the sample down to 5 K ~\cite{kummer}. A bulk polycrystalline sample (2$\times$2$\times$5 mm$^3$) was cleaved in an ultra-high vacuum chamber (base pressure of the order of $10^{-9}$ mbar) before being transferred to the superconducting magnet. XAS was detected in total-electron-yield (TEY) mode. XMCD spectra at the Cr L$_{2,3}$ and Fe L$_{2,3}$ edges were obtained as the difference between XAS spectra measured with opposite helicities ( $\mu_{+}$ and $\mu_{-} )$ in a finite magnetic field. XAS and XMCD measurements were also performed at the BL14 beamline of the HiSOR synchrotron radiation center (Hiroshima, Japan).

\begin{table}
\caption{Atomic positions (top) and lattice parameters (bottom) for Cr$_{2-x}$Fe$_x$GeC with $x =$ 0.00, 0.05 and 0.1.}
\centering
\setlength{\tabcolsep}{12pt}
\begin{tabular}{ccccc}
\hline 
\hline
Sites & Wyckoff symbol&  x & y  & z \\ 
\hline													
Cr/Fe & 4$f$&0.333 & 0.666 & 0.586  \\ 
Ge & 2$c$&0.333 & 0.666 & 0.25  \\ 				
C & 2$a$&0 & 0 & 0  \\ 
\hline
\hline
\end{tabular}
\vskip 0.5cm
\begin{tabular}{ccc}
\hline 
\hline
Sample & $a$ (\AA) & $c$ (\AA)  \\ 
\hline													
$x=$ 0.00 & 2.954 & 12.107  \\ 
$x=$ 0.05 & 2.955 & 12.108   \\ 				
$x=$ 0.10 & 2.956 & 12.112  \\ 	
\hline
\hline 
\end{tabular}
\label{tab:lat}
\end{table} 
\begin{figure}[htbp]
	\centering
	\includegraphics[width = 8 cm]{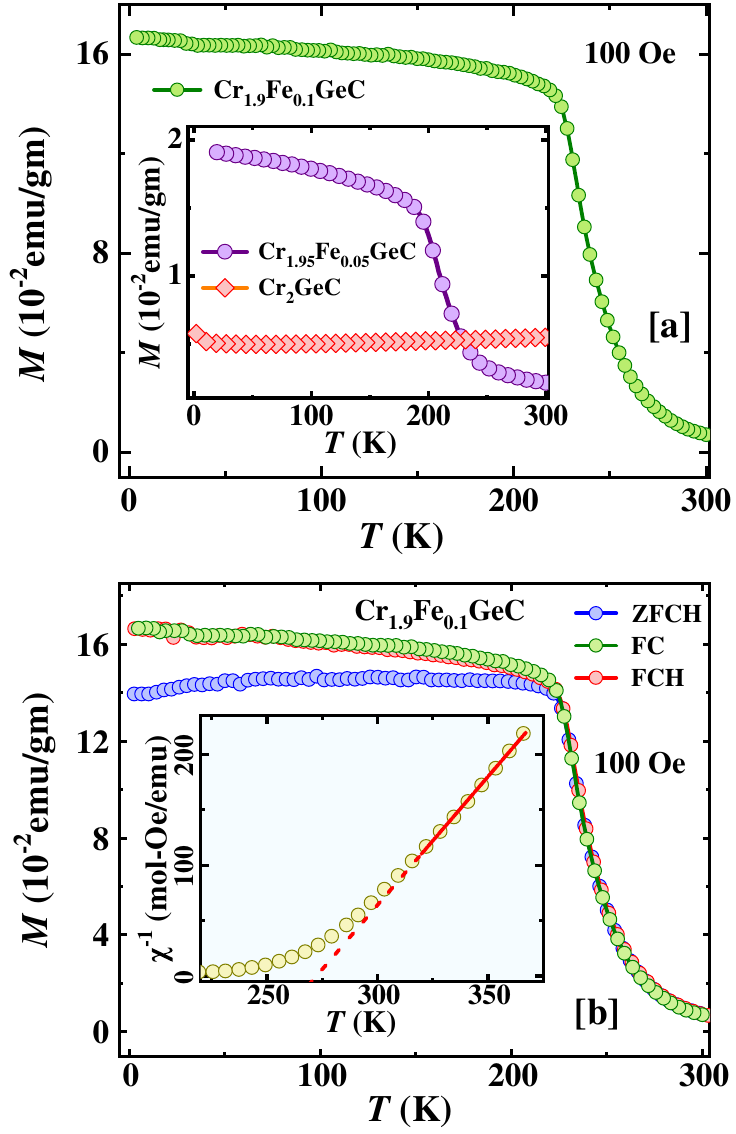}
	\caption {(a) shows the  $T$ variation of $M$ for $x =$ 0.1 sample measured at 100 Oe during cooling. Inset compares $M$-$T$ curves for $x =$ 0.00 and  0.05 samples. (b) depicts $M$-$T$ curves with zero field-cooled-heating (ZFCH), field cooled (FC) and field-cooled-heating (FCH) protocols, measured under $H$ = 100 Oe for $x =$ 0.1. The inset of (b) shows the Curie Weiss fit (solid line) to the inverse $\chi$ vs $T$ plot at high temperature region.}
	\label{fig:mt}
\end{figure}

\section{Results and Discussion}
\subsection{Magnetization}
 Fig.~\ref{fig:mt} [a] and its inset depict the $T$ variation of $M$ of three samples Cr$_2$GeC, Cr$_{1.95}$Fe$_{0.05}$GeC and Cr$_{1.9}$Fe$_{0.1}$GeC under $H$ = 100 Oe in the field-cooled heating protocol. $M$ is found to be small and positive for the parent sample $x=$ 0 with an insignificant $T$-dependence, which corroborates the previous report of the Pauli paramagnetic nature of the sample~\cite{Liu2014,Lin2016}. $x=$ 0.05 and 0.1 samples order magnetically below room temperature, as evident from the rising features in the $M$ vs $T$ cooling data (see Fig.~\ref{fig:mt} [a]). The critical points for magnetic ordering for the $x=$ 0.05 and 0.1 samples are found to be 207 K and 232 K, respectively, from the $dM/dT$ vs. $T$ plot (not shown here).
\par
Fig.~\ref{fig:mt} [b] shows the $M$ versus $T$ data of $x =$ 0.1 in the zero-field-cooled (ZFCH), field-cooling (FC) and field-cooled-heating (FCH) modes. ZFCH and FC data separate from each other at around 225 K. The FC and FCH data do not show any thermal hysteresis, indicating that the magnetic transition is second order in nature.
\par
High-temperature magnetic susceptibility ($\chi=M/H$) have been fitted with a Curie-Weiss law (straight line fit to the $\chi^{-1}$ vs. $T$ plot~\cite{mondal}, depicted for $x=$ 0.1 sample in the inset of Fig.~\ref{fig:mt} [b]) for $x=$ 0.05 and 0.1 samples. From the Curie-Weiss fitting, the effective paramagnetic moments per formula unit (f.u.) and Curie-Weiss temperatures are found to be $\mu_{eff}$ = 1.85 $\mu_B$(f.u.)$^{-1}$ and $\theta_p=$ 272 K, respectively for Cr$_{1.9}$Fe$_{0.1}$GeC composition. For $x =$ 0.05 sample, the Curie-Weiss fitting provides $\mu_{eff}=$ 1.54 $\mu_B$(f.u.)$^{-1}$ and $\theta_p=$ 225 K. Positive $\theta_p$ indicates ferromagnetic interaction in both compositions. In particular, $\theta_p$ is higher for the highest Fe-doped sample, indicating an increase in magnetic correlation with Fe content.

\begin{figure}[htbp]
	\centering
	\includegraphics[width = 8.22 cm]{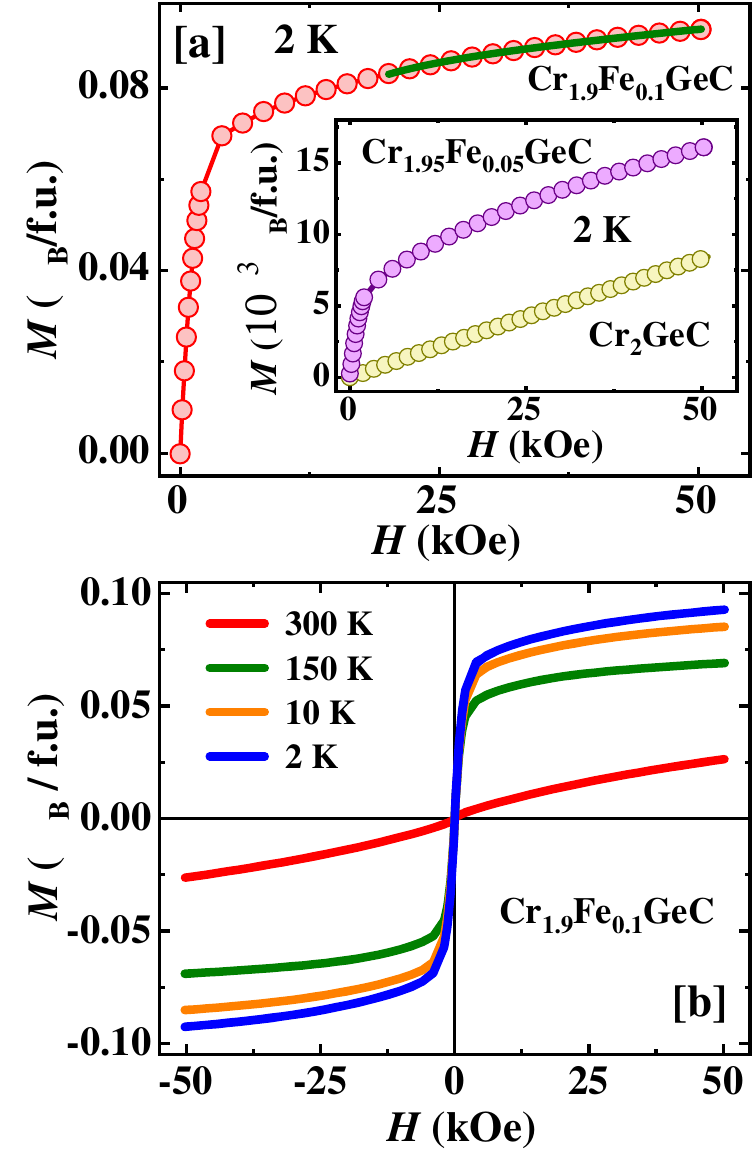}
	\caption {(a) shows the $M$-$H$ data measured at $T =$ 2 K for $x =$  0.1 sample. Solid olive line represents fitting to the high field $M$-$H$ data with equation $M = q_s(1-\zeta/H^2) +\chi_HH$. Inset of (a) compares $M$-$H$ curves for $x$ = 0.00 and $x$ = 0.05 samples. (b) depicts $M$-$H$ curves measured at different temperatures for Cr$_{1.9}$Fe$_{0.1}$GeC.}
	\label{fig:mh}
\end{figure}
\par	
 Fig.~\ref{fig:mh} [a] and its inset show the isothermal $M$ versus $H$ data recorded at 2 K for three compounds with a maximum applied magnetic field of 50 kOe. The parent sample shows a linear variation of $M$ with $H$, which is expected for a Pauli paramagnet. For doped samples, the $M$-$H$ curves have a typical FM nature with a sharp increase at low fields followed by a tendency toward saturation at higher fields. In particular, the isotherms for the doped samples do not completely saturate at the maximum applied field of 50 kOe. However, the value of $M$ at 50 kOe  is found to be higher for the $x=$ 0.1 sample. Fig.~\ref{fig:mh} [b] shows few isotherms at different temperatures for Cr$_{1.9}$Fe$_{0.1}$GeC sample. The moment observed at 2 K for 50 kOe of applied field is found to be around $\sim$ 0.08 $\mu_B$(f.u.)$^{-1}$ indicating the weak FM character of the sample. The coercive field of the sample is rather small, indicating a very soft FM nature. Even at 300 K, the $M$ versus $H$ curve is not a straight line, indicating that the sample has some short-range correlations at room temperature.
\par
Itinerant character of a ferromagnet can be understood by studying the Rhodes-Wolhfarth ratio ~\cite{Matthias,santiago,rhodes} RWR = $q_c/q_s$. The saturation moment $q_s$ is inferred from the low-temperature $M$-$H$ data, and  $q_c$ is related to the effective paramagnetic moment (obtained from the high-$T$ Curie-Weiss fitting): $\mu_{eff}^{2} = q_c(q_c+2)$. Since $\mu_{eff} =$ 1.85 $\mu_B$(f.u.)$^{-1}$ for $x =$ 0.1 sample, $q_c$ turns out to be 1.103 $\mu_B$(f.u.)$^{-1}$.  Fe doped Cr$_2$GeC does not show complete saturation, and we have used the `law of approach to saturation magnetization', $M= q_s(1-\zeta/H^2) +\chi_HH$, to fit the high-field $M$-$H$ isotherm for 2 K to obtain $q_s$ (olive solid line in Fig.~\ref{fig:mh} [a]). From fitting $q_s$ = 0.083 $\mu_B$(f.u.)$^{-1}$, $\zeta$ = 21.67 $\mu_B$-Oe$^2$(f.u.)$^{-1}$ and $\chi_H$ = 2$\times$10$^{-4} \mu_B$(Oe-f.u.)$^{-1}$. Using these values of $q_c$ and $q_s$, the resulting RWR for the compound Cr$_{1.9}$Fe$_{0.1}$GeC is found to be 13.29.  This value of RWR is much higher than unity and is comparable to other WIFMs (see Table~\ref{WIFM}) such as ZrZn$_2$, Y$_4$Co$_3$. 

\begin{table*}
 \caption{Some basic parameters of Cr$_{1.9}$Fe$_{0.1}$GeC and few other itinerant weak ferromagnets.}
 \setlength{\tabcolsep}{0.3cm}
 \vskip 0.15cm
 \centering
 \begin{tabular}{c c c c c c c c}
   \hline
   \hline
     Sample & $\theta_p$ (K) &$T_c$ (K) & q$_c$ [$\mu_B$(f.u.)$^{-1}$]& q$_s$ [$\mu_B$(f.u.)$^{-1}$] & RWR & $\Gamma$ (mJmol$^{-1}$K$^{-2}$) & References\\ 
   \hline													
    Cr$_{1.9}$Fe$_{0.1}$GeC&272 & 246 & 1.103&0.083&13.29&27&this work  \\ 
    Y$_2$Ni$_7$ &40 & 53 &0.37&0.06&6.17&52.3 &~\cite{Bhattacharyya}   \\ 						
    ZrZn$_2$ &33 & 21 & 0.65 & 0.12 & 5.4 & 45 &~\cite{Pickart,Yelland}  \\ 
    InSc$_3$ &8 & 6& 0.26&0.045&5.75&12 &~\cite{Matthias,Aguayo}  \\ 
    Y$_4$Co$_3$ &14 &5 & 0.14&0.012&11.5&3.5 &~\cite{Takigawa}  \\ 	
    Co$_3$SnC &4.6 &3.6 & 0.45&0.09&5.17&36.5 &~\cite{wang}  \\ 
   \hline 
   \hline
 \end{tabular}
 \label{WIFM}
 \end{table*}  
 
\begin{figure*}[htbp]
        \centering
	\includegraphics[width = 18 cm]{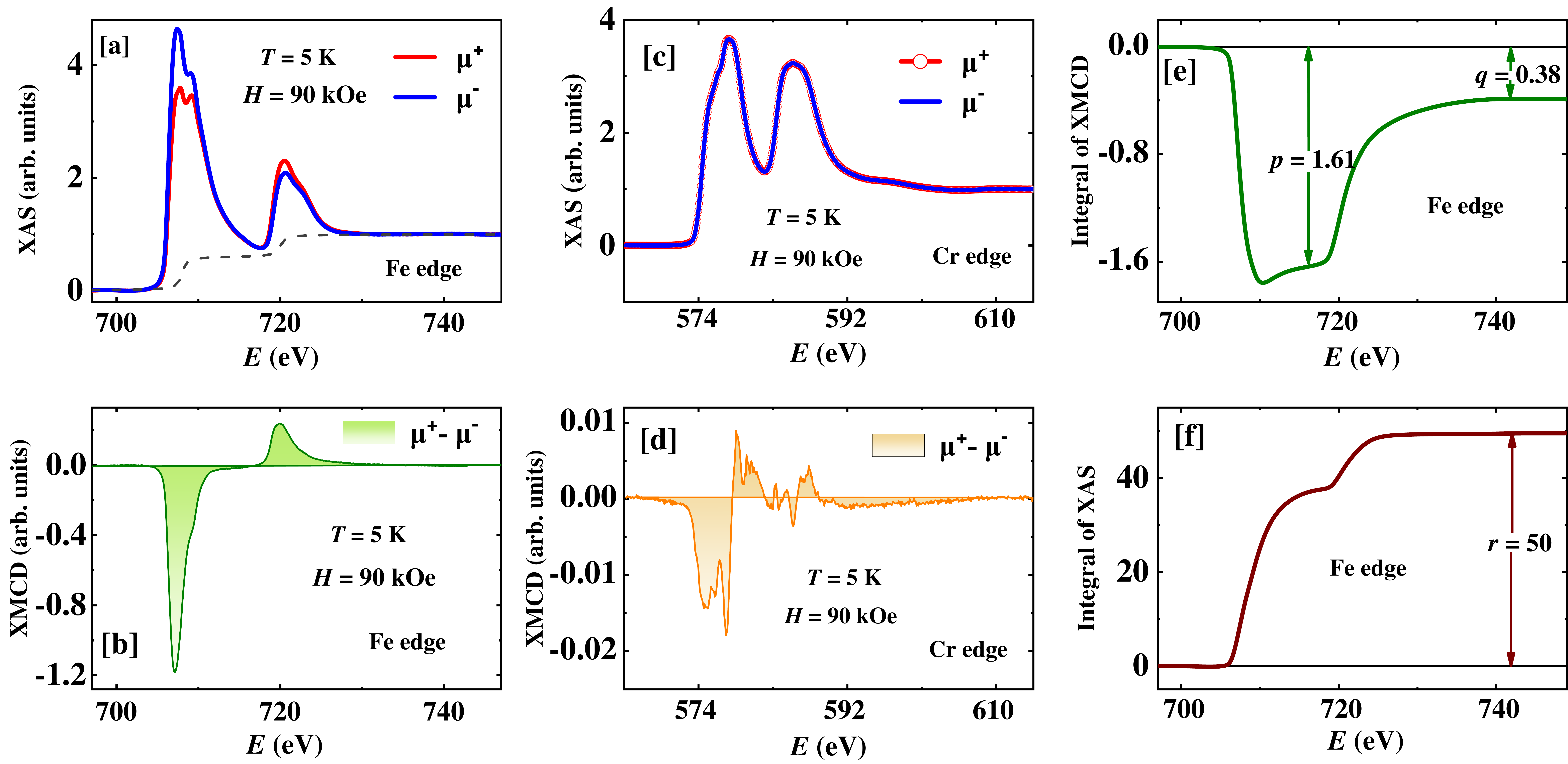}
	\caption {(a) XAS for left circular ($\mu_{-}$) polarized light, right circular ($\mu_{+}$) polarized light measured in  TEY mode at $T$ = 5 K under $H$= 90 kOe of applied field at Fe L$_{2,3}$ for Cr$_{1.9}$Fe$_{0.1}$GeC sample. Black dashed line represents the two-step background function. (b) depicts the corresponding XMCD signal from the Fe-edge.  (c) and (d) respectively represent the XAS and XMCD signals from the Cr L$_{2,3}$ edges. (e) and (f)  respectively show the energy variation of the integrals over the XMCD ($p$ and $q$) and  XAS ($r$) signals. $p$, $q$, and $r$ are defined in the text.}
	\label{fig:xmcd}
\end{figure*}

\subsection{X-ray magnetic circular dichroism (XMCD)}

\par
Fig.~\ref{fig:xmcd} [a] and [c] show the XAS spectra for Cr$_{1.9}$Fe$_{0.1}$GeC measured in TEY mode using both left ($\mu^{+}$) and right ($\mu^{-}$) circularly polarized x-rays at Cr L$_{2,3}$ and Fe L$_{2,3}$ edges at 5 K under an applied field of 90 kOe. The Cr L$_{2,3}$ edges XAS spectra exhibit two broad spin-orbit split peaks: the L$_3$ peak at 578 eV ($2p_{3/2}$) and the L$_2$ peak at 586 eV ($2p_{1/2}$), which result from dipole-allowed transitions from the 2$p$ core to the unoccupied 3$d$ states. Similarly, the Fe L$_{2,3}$ edge XAS spectra display two peaks at 707 eV ($2p_{3/2}$) and 720 eV ($2p_{1/2}$).

\par
The XMCD signal is defined as $\Delta\mu = \mu^+ - \mu^-$, where $\mu^+$ and $\mu^-$ denote the L$_{2,3}$ XAS spectra for photon helicity parallel and antiparallel to the magnetization direction, respectively. A strong XMCD signal is observed for the Fe element (see Fig.~\ref{fig:xmcd} [b]), while the Cr signal is vanishingly small (see Fig.~\ref{fig:xmcd} [d], almost two orders of magnitude smaller). Since XMCD provides an element-specific measurement of magnetization, it allows for the determination of the orbital ($\langle L_Z \rangle$) and effective spin ($\langle S_Z \rangle$) moments using the magneto-optical sum rules~\cite{Chen1995}. The energy-dependent variation of the XMCD data enables the calculation of these moments in terms of the following integrals:

\begin{align}
	\langle L_Z \rangle &= \frac{-4n_h \displaystyle\int_{{\rm L_3+L_2}} \Delta\mu(\omega) d\omega}{3\displaystyle\int_{{\rm L_3+L_2}}(\mu^+ + \mu^-) d\omega} \nonumber \\[10pt]
	&= \frac{-2q n_h}{3r}
\end{align}

\begin{align} 
	\langle S_Z \rangle &= \frac{n_h \left[ 4\displaystyle\int_{{\rm L_3+L_2}} \Delta\mu(\omega) d\omega - 6\displaystyle\int_{{\rm L_3}} \Delta\mu(\omega) d\omega \right ]}{\displaystyle\int_{{\rm L_3+L_2}}(\mu^+ + \mu^-) d\omega} \nonumber \\[10pt]
	&= \frac{(2q -3p)n_h}{r}
\end{align}

Here, $n_h$ represents the number of 3$d$ holes. The quantities $p$, $q$, and $r$ are defined as follows:
$p = \int_{{\rm L_3}} \Delta\mu(\omega) d\omega$ represents the integral of the XMCD signal over the L$_3$ edge,  
$q = \int_{{\rm L_3+L_2}} \Delta\mu(\omega) d\omega$ is the integral on both the edges L$_3$ and L$_2$,  
and $r = \frac{1}{2}\int_{{\rm L_3+L_2}} (\mu^+ + \mu^-) d\omega$ denotes the integrated area of the total XAS signal over both edges after subtracting a background modeled by step functions. The integration ranges are 707–715 eV for the L$_3$ edge and 707–750 eV for the combined L$_2$ and L$_3$ edges. At 5 K, the calculated values of $\langle L_Z \rangle$ and $\langle S_Z \rangle$ are 0.034 $\mu_B$/Fe and 0.6 $\mu_B$/Fe, respectively. The orbital moment of the Fe atom is significantly smaller than the spin moment, indicating a system with very weak spin-orbit coupling. This is expected for a 3$d$ element such as Fe, where the orbital moment is quenched because of strong crystal field effects. From the field-dependent bulk macroscopic magnetization measurements at 2 K, the saturation moment is approximately 0.083 $\mu_B$(f.u.)$^{-1}$ or 0.83 $\mu_B$/Fe (considering the Fe stoichiometry to be 0.1 per formula unit). This value is in close agreement with the microscopic XMCD result. We also estimated spin and orbital moments at the Cr-L edges. The reliable application of the XMCD sum rules at the Cr-L edges is usually prohibited by the too small 2$p$ spin-orbit splitting causing an overlap of the Cr $L_3$ and $L_2$ edges. We can however give as upper estimates for $\langle L_Z \rangle =$ 0.002 $\mu_B$ and, less accurately, $\langle S_Z \rangle =$ 0.006 $\mu_B$/Cr.

\par
Pathirage \textit{et al.} recently studied Cr-intercalated VS$_2$ dichalcogenides by XMCD measurements, and a significant moment was detected at the  Cr-edge~\cite{khatun}. XMCD studies on some layered van der Waals magnetic materials such as CrTe$_2$, Fe$_5$GeTe$_2$, and Cr$_2$Ge$_2$Te$_6$ reveal that the magnetic moment arises primarily from spin contributions rather than orbital contributions~\cite{suzuki,zhang,yamagami}. Consistent with our findings, Mijit \textit{et al.} confirmed the WIFM character of CoS$_2$ through XMCD measurements at the Co-L edge~\cite{mijit}. The spin moment at the Co site via L-edge XMCD is found to be 0.78 $\mu_B$/Co, which is comparable to the Fe moment observed in our sample.

\par
\begin{figure}[htbp]
	\centering
	\includegraphics[width = 8 cm]{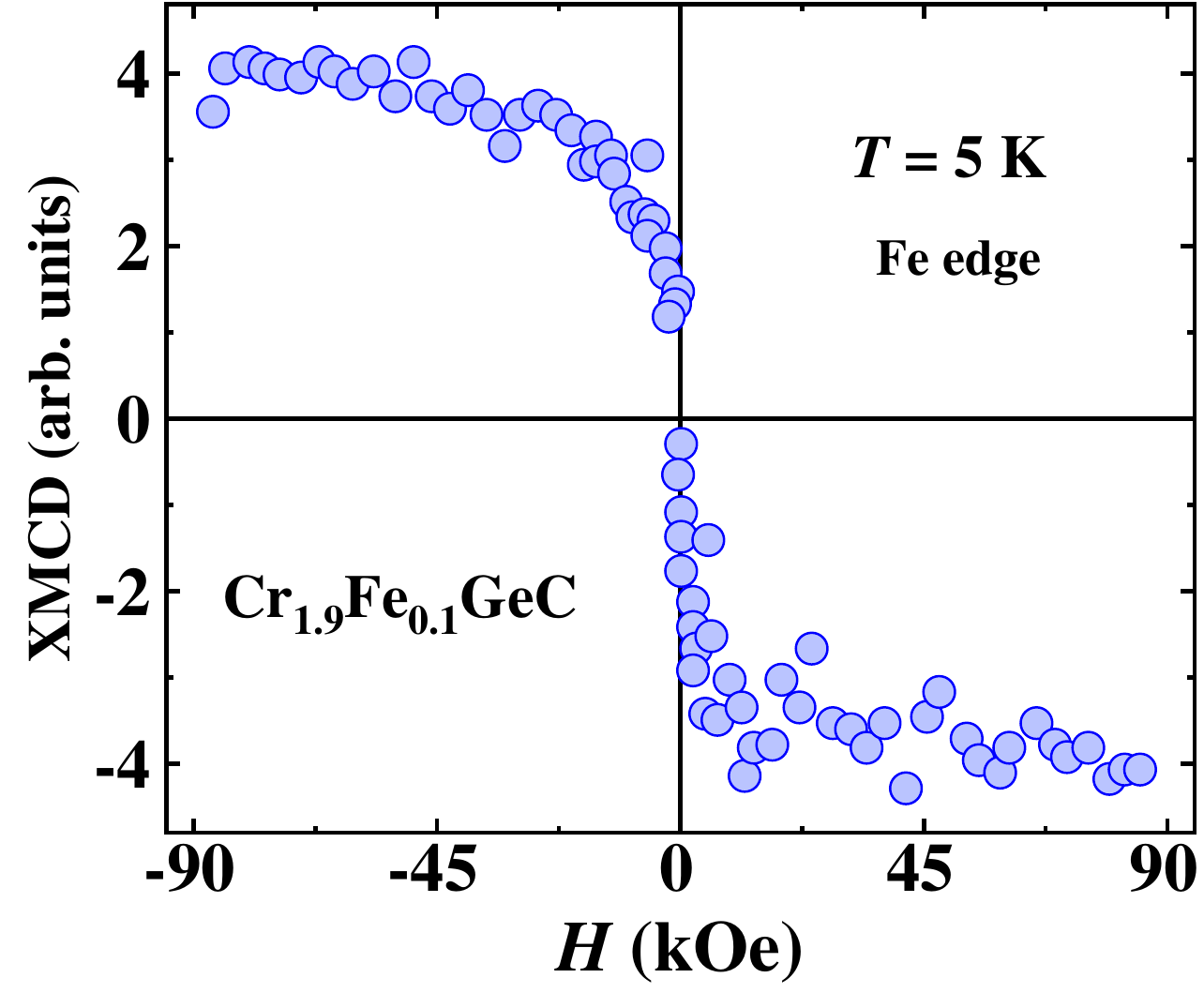}
	\caption {Field variation of XMCD signal of the Fe-$L_3$ edge at 5 K for the compound Cr$_{1.9}$Fe$_{0.1}$GeC.}
	\label{fig:xmcd_H}
\end{figure}

From the field variation of the element-specific XMCD signal, we can gain insight into the magnetic properties of the system~\cite{i-max,yamamoto,suwa,chen}. The XMCD signal from the Fe-L$_3$ edge was recorded as a function of the applied field, and we have plotted the area under the XMCD signal [$p(H) =\int_{{\rm L_3}} \Delta\mu(\omega) d\omega$] with $H$ in Fig.~\ref{fig:xmcd_H}. The field-dependent XMCD signal mimics the macroscopic $M$ versus $H$ except a change in sign. In our case, the XMCD signal tends to saturate at low magnetic fields, similar to the $M$-$H$ curve.
\begin{table*}
 \caption{Spin and orbital moments obtained from XMCD measurement for some transition metal based alloys and compounds.}
 \setlength{\tabcolsep}{0.3cm}
 \vskip 0.15cm
 \centering
 \begin{tabular}{c c c c c c c c}
   \hline
   \hline
    Sample & atom & edge & $\langle L_Z \rangle$ ($\mu_B$/atom) & $\langle S_Z \rangle$ ($\mu_B$/atom)& $\langle S_Z \rangle$/$\langle L_Z \rangle$ & References \\ 
   \hline													
    Cr$_{1.9}$Fe$_{0.1}$GeC & Cr, Fe & $L$ & 0.002, 0.034 & 0.006, 0.6 & 0.33, 17.65 & this work  \\
    Cr$_x$VS$_2$ & V, Cr & $L$ & - & 0.058,1.4 & - & ~\cite{khatun}  \\
    Cr$_2$Ge$_2$Te$_6$ & Cr & $L$ & 0.059 & 1.92 & 32.54 & ~\cite{suzuki}  \\
    CrTe$_2$ & Cr & $L$ & 0.08 & 2.85 & 35.63& ~\cite{zhang}  \\
    Fe$_5$GeTe$_2$ & Fe & $L$ & 0.1 & 1.8 & 18 & ~\cite{yamagami}  \\
    CoS$_2$ & Co & $L$ & 0.049 & 0.782 & 15.96 & ~\cite{mijit}  \\
   \hline 
   \hline
 \end{tabular}
 \label{xas}
 \end{table*}

\subsection{Electrical Resistivity}
\begin{figure}[htbp]
	\centering
	\includegraphics[width = 9 cm]{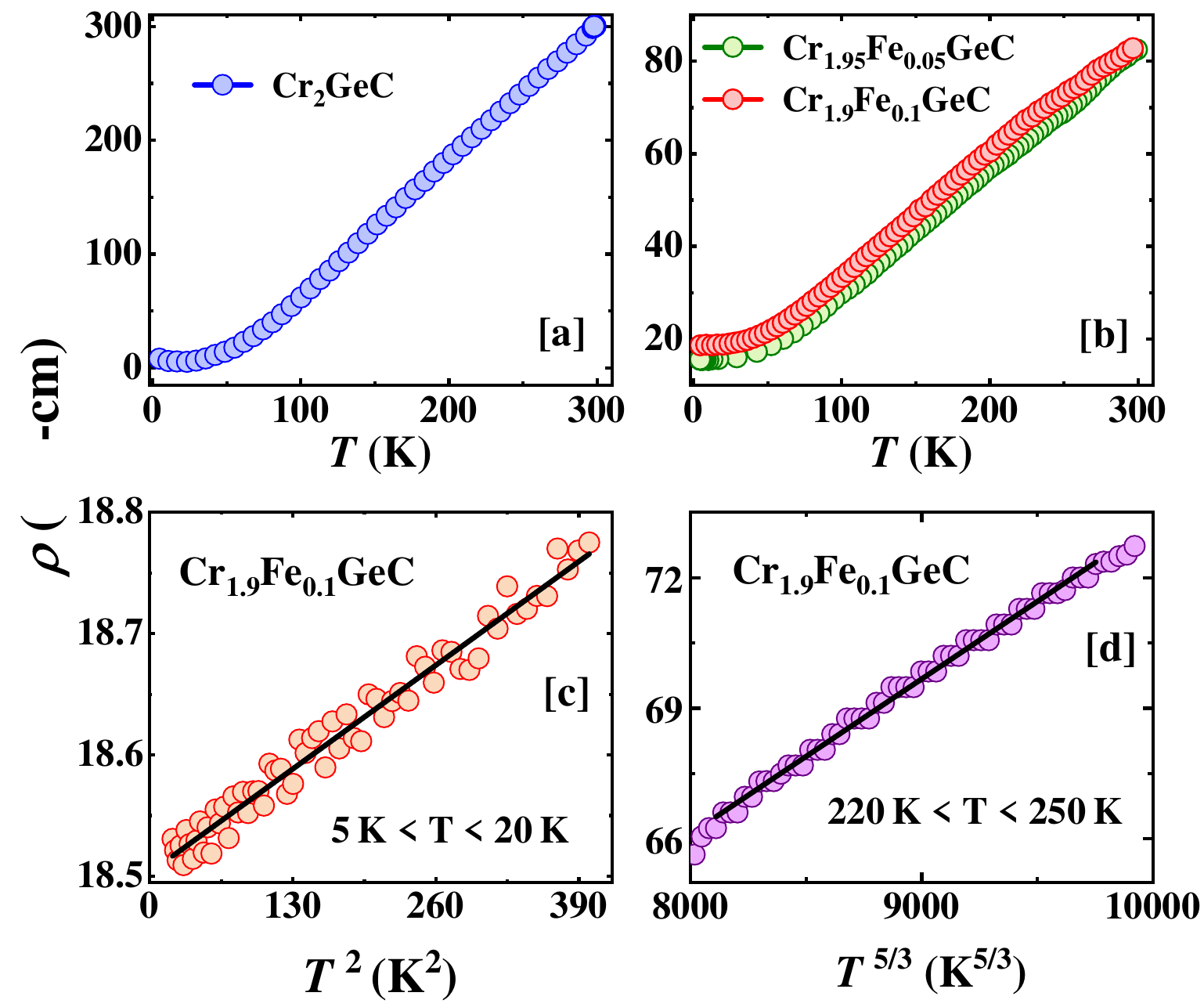}
	\caption{(a) depicts the temperature variation of resistivity for Cr$_2$GeC sample, while the $\rho$ versus $T$ plots for  $x =$ 0.05 and $x =$  0.1 are shown in (b). (c) and (d) represent the $\rho$ vs. $T^2$ plot (5 K$ <T< $20 K) and $\rho$ vs. $T^{5/3}$ plot (220 K$<T<$250 K) for $x =$  0.1 sample, respectively. The solid lines though the data points are linear fit to the curves.}
\label{fig:RT}
\end{figure}

The zero-field $\rho$ versus $T$ data, shown in Figs.~\ref{fig:RT} [a] and [b], indicate a clear metallic nature for the three samples in the temperature range of 2 to 300 K. With Fe doping, $\rho$ decreases compared to the parent sample. The residual resistivity ratio [RRR = $\rho(300K)/ \rho(5K)$] is 45, 5 and 4.5 for Cr$_2$GeC, Cr$_{1.95}$Fe$_{0.05}$GeC and Cr$_{1.9}$Fe$_{0.1}$GeC, respectively. The high value of residual resistivity in the parent sample indicates that it is a high-quality sample.
\par
$\rho(T)$ for the $x=$ 0.1 sample (Fig.~\ref{fig:RT} [c]) exhibits a well defined $T^2$ dependence ($\rho = \rho_0 +  {\rho_t}T^2$ ) at low $T$ (5 K $< T <$ 20 K), which is the usual Fermi liquid behavior in common metals resulting from electron electron scattering. The SCR theory for WIFM predicts a term $T^2$ in $\rho$, which arises from the scattering of conduction electrons with fluctuating moments~\cite{Ueda, Hertel}. In the case of Cr$_{1.9}$Fe$_{0.1}$GeC, the coefficient of the term $T^2$ is notably significant, measuring 5 $\times$ $10^{-10}  \Omega$-cm K$^{-2}$, which is approximately an order of magnitude higher than that of typical FM metals such as Ni and Fe ($\sim 10^{-11}  \Omega$-cm K$^{-2}$). Similarly, elevated $\rho_t$ values were observed in prototype WIFMs, such as  ZrZn$_2$ ~\cite{Ogawa} or Ni$_3$Al~\cite{Ogawa2}, and it is generally thought to be associated with spin fluctuations.

\par
According to the SCR theory, $\rho$ should vary with $T^{5/3}$ just below the magnetic ordering temperature for WIFM~\cite{Ueda}. Fig.~\ref{fig:RT} [d] shows $\rho$ as a function of $T^{5/3}$ between 220 K and 250 K and the linear nature of the curve indicates that $\rho$ varies as $T^{5/3}$, which is consistent with the prediction of the SCR model.

 \begin{figure}[htbp]
 	\centering
 	\includegraphics[width = 8 cm]{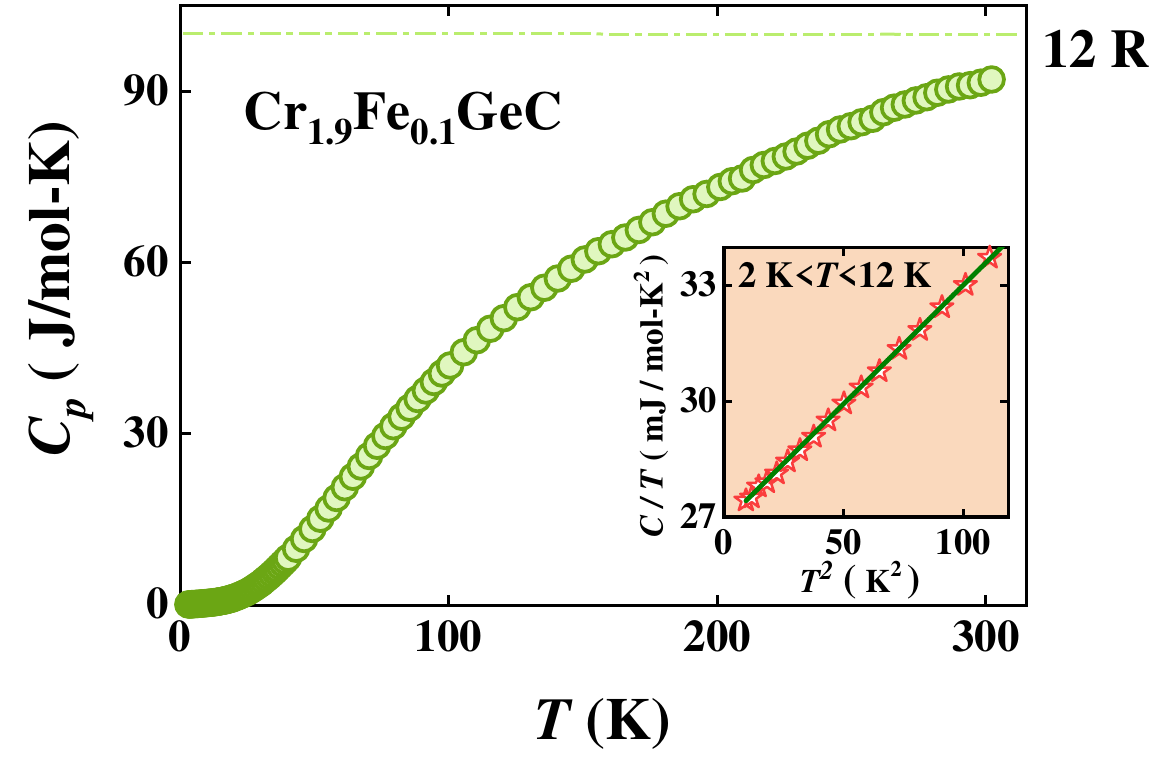}
 	\caption {Temperature variation of Specific heat for Cr$_{1.9}$Fe$_{0.1}$GeC sample. Inset depicts the low temperature linear fit to the $C/T$ vs $T^2$ plot.}
 	\label{fig:cp}
 \end{figure}

\subsection{Heat Capacity}
In Fig.~\ref{fig:cp}, the $T$ variation  of the specific heat ($C_p$) is shown in the range of 2-300 K for $x=$ 0.10. No anomaly is observed around the magnetic transition. This is possibly due to the low moment and the relatively high value of the transition temperature, with the magnetic contribution being much smaller than the phonon contribution toward $C_p$. The low-$T$ $C_p/T$ vs. $T^2$ plot (inset of Fig.~\ref{fig:cp}) shows a linear variation for $T\leq$ 12 K. For $T \ll \Theta_D$ (= Debye temperature), the heat capacity varies as $C/T = \Gamma + BT^2$, where $\Gamma$ and $B$ are the coefficients of electronic and lattice contributions of $C_p$, respectively. The solid line in the inset of Fig.~\ref{fig:cp} represents a fit to the data, and we obtain $\Gamma$ = 27 mJ mol$^{-1}$K$^{-2}$ and $B$ = 0.061 mJ mol$^{-1}$K$^{-4}$. The value of $\Theta_D$ obtained from $B$ is 503 K. The value of the electronic specific heat coefficient $\Gamma$ is comparable to many other WIFM systems (see Table~\ref{WIFM}).     

\par
 The obtained enhanced values of $\rho_t$ and $\Gamma$ from the $T$ variation of electrical  resistivity and the specific heat, respectively, indicate typical Fermi liquid-like behavior. The quantity $\rho_t/\Gamma^2$, known as the Kadowaki-Woods ratio~\cite{okabe}, is an important parameter to determine Fermi-liquid state in a metal. It is known that $\rho_t\propto m^{*2}$ and $\Gamma \propto m^*$, where $m^*$ is the effective mass of conduction electrons. Therefore, within a class of materials that obey the renormalized band picture, the ratio $\rho_t/\Gamma^2$ should have a universal value. For heavy-fermion metals, the ratio is found to be close to 1.0 $\times$ $10^{-5}$ $\Omega$-cm-mol$^2$ K$^2$J$^{-2}$~\cite{kadowaki}. However, for transition metals, the ratio has an average value of $\sim$ 10$^{-6}$ $\Omega$-cm-mol$^2$ K$^2$J$^{-2}$, which is one order of magnitude lower than that of heavy fermions. We have calculated the ratio for Cr$_{1.9}$Fe$_{0.1}$GeC and it turns out to be 6.5 $\times$ 10$^{-6}$ $\Omega$-cm-mol$^2$ K$^2$J$^{-2}$, which is fairly close to the value found in the case of transition metals~\cite{Bhattacharyya,wang}. As $\rho_t$ and $\Gamma$ both increase compared to the normal metal, the Kadowaki-Woods ratio remains unchanged. This indicates that the spin fluctuations in Cr$_{1.9}$Fe$_{0.1}$GeC can be well accounted for by the renormalized electronic band parameters. 

\begin{figure}[htbp]
\centering
\includegraphics[width = 8 cm]{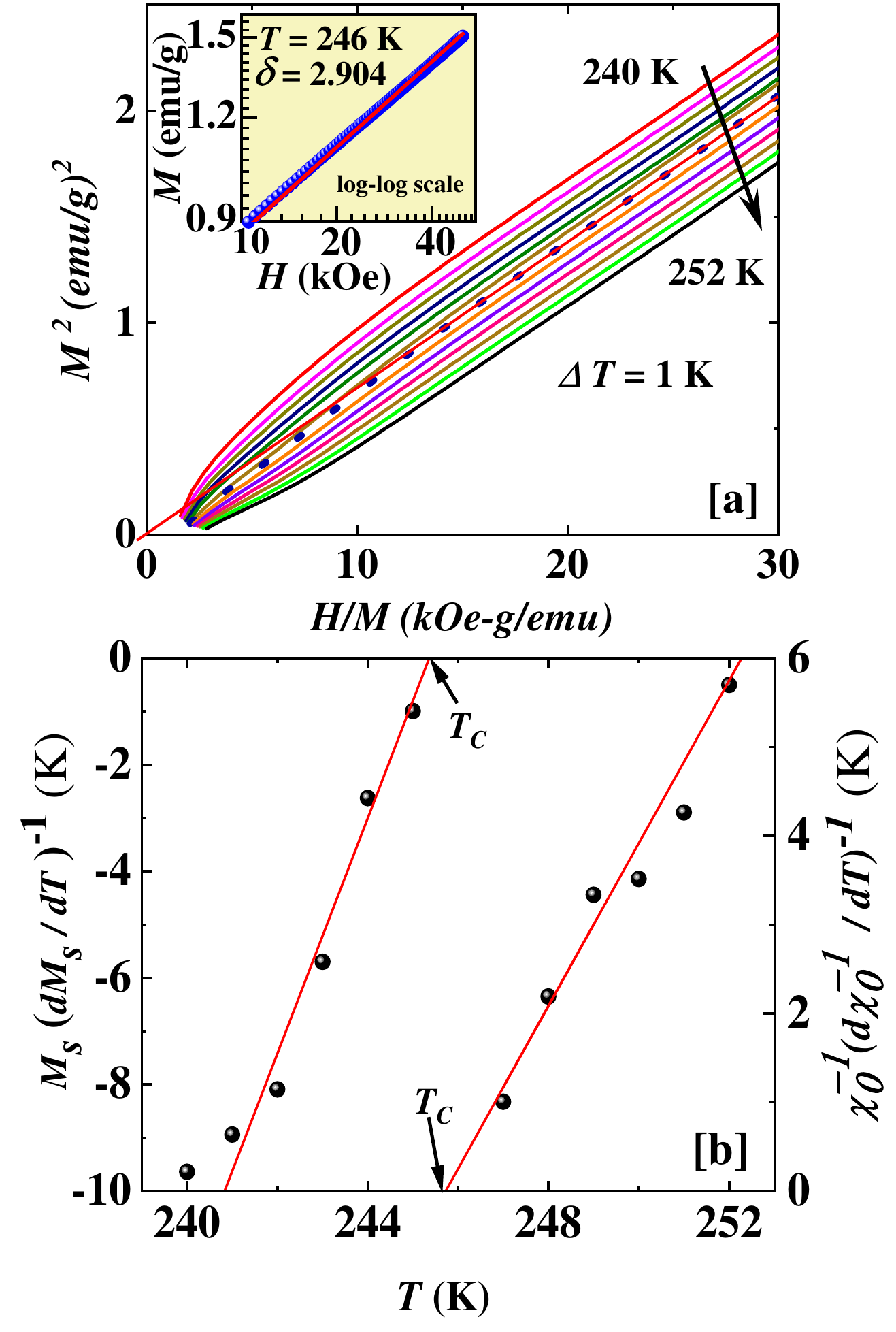}
\caption {(a) shows several $M^2$ vs $(H/M)$ isotherms (Arrott plot) of Cr$_{1.9}$Fe$_{0.1}$GeC around $T_C$  with temperature interval $\Delta T$ = 1 K. The inset shows the $M$-$H$ plot in the log-log scale at the  critical temperature $T_C =$ 246 K along with a straight line fit to the data.  (b) Kouvel-Fisher plot of  $M_S$ (left axis) and  $\chi^{-1}_{0}$ (right axis) for the sample Cr$_{1.9}$Fe$_{0.1}$GeC. Straight lines are the  linear fit to the data.}
\label{Arott}
\end{figure}

\subsection{Critical behavior}
\subsubsection{Arrot plot}
The standard method for determining the critical exponents is the use of the Arrot plot. To construct the $M^2$ vs $H/M$ Arott plot for Cr$_{1.9}$Fe$_{0.1}$GeC, several isotherms were recorded around the Curie point with temperature separation $\Delta T$ = 1 K (see Fig.~\ref{Arott} [a]). All the $M^2$ vs $H/M$ isotherms show a positive slope indicating the magnetic transition to be second order in nature~\cite{Banerjee}. We observe that the high-field part of the Arrot curves are a set of parallel straight lines, and an extrapolation of the high-field part of the isotherm recorded at 246 K passes through the origin. It indicates that the Curie point of Cr$_{1.9}$Fe$_{0.1}$GeC is $T_C =$ 246 K, which is slightly higher than the value obtained (232 K) from the curve $dM/dT$ vs. $T$ plot.
\par
The high-field linear nature of the Arrot plot indicates that the system follows the mean-field theory. Near a second-order phase transition, the diverging correlation length leads to the universal scaling laws for spontaneous magnetization ($M_S$) below $T_C$, initial susceptibility ($\chi_0 = \lim_{H \to 0} M/H$) above $T_C$, and magnetization at $T_C$ via a set of critical exponents $\beta$, $\gamma$, and $\delta$, which are defined as~\cite{Stanley} 
\begin{eqnarray}
\label{crit}
M_S(T)&=&M_0 {|\epsilon|}^{\beta},  \epsilon < 0   \nonumber \\
\chi^{-1}_{0} (T)& =& G(\epsilon)^{\gamma} , \epsilon > 0   \nonumber \\ 
M &=& XH^{1/\delta} ,  \epsilon = 0   
\end{eqnarray}
Where $\epsilon$ = $(T-T_C)/T_C$ is the reduced temperature. $M_0$, $G$ and $X$ are the critical amplitudes.
\par
Since, our Arrot plots are linear, we can assume the critical exponents have the mean field values, namely $\beta =$ 0.5, $\gamma =$ 1, and $\delta =$ 3.  

\subsubsection{Kouvel-Fisher plot}
To check the authenticity of the critical analysis, the Kouvel-Fisher (KF) method can be used. The KF plot is based on the following equations~\cite{Kouvel}:
\begin{eqnarray}
\frac{M_S}{dM_S/dT}  = \frac{T-T_C}{\beta} \nonumber \\
\frac{\chi_0^{-1}}{d\chi_0^{-1}/dT} =\frac{T-T_C}{\gamma}
\label{kf1}
\end{eqnarray}
According to  the KF equation, the $T$-dependence of $\frac{M_S}{dM_S/dT}$ and $\frac{\chi_0^{-1}}{d\chi_0^{-1}/dT}$ should be straight lines with slopes 1/$\beta$ and 1/$\gamma$ respectively, and their intercept on the $T$-axis will result $T_C$. The critical exponent $\delta$ can be obtained by Widom scaling relation,
\begin{equation}
\delta = 1+\gamma/\beta
\label{widom}
\end{equation}

We have extrapolated the high-field linear part of the $M^2$ versus $H/M$  Arrot plot, and $M_S^2$ and $\chi_0 ^{-1}$ are obtained as the intercepts on the vertical and horizontal axes, respectively. Using the values of $M_S^2$ and $\chi_0 ^{-1}$, we have constructed the KF plots as shown in Fig.~\ref{Arott} [b]. From the slope of the curves we obtain $\beta$ = 0.498, $\gamma$ = 1.05, which are quite close to the mean field values. The critical point is found to be $T_C$ = 245.5 K. Using the Widom scaling relation, we obtain $\delta =$ 3.11.   

\par
From eqn.~\ref{crit}, it is evident that the slope of the $M$ vs. $H$ curve in the $\log$-$\log$ scale at $T_C$ is equal to $\delta^{-1}$. The slope is found to be 0.34(4)  (see inset of Fig. ~\ref{Arott} [a]) indicating $\delta =$ 2.904 , which is in good agreement with the value of $\delta$  obtained from the Widom scaling relation.

\section{Summary and conclusion}
 The MAX phase compound Cr$_2$GeC is a Pauli paramagnet without showing any localized moment. The present investigation indicates that small Fe doping at the Cr site (2.5 \%) induces ferromagnetism. The induced moment in the FM state is rather low, and the Fe-doped Cr$_2$GeC can be identified as a weak itinerant ferromagnetic system. The WIFM character is further evident from the large value of Rhodes-Wolhfarth ratio. We also observe a relatively large value of the electronic specific heat and the electron-electron scattering term of the resistivity, which indicate large spin-fluctuation akin to the other itinerant magnets.

\par
Interestingly, our XMCD investigation indicates that the magnetic moment in the doped samples arises solely from Fe atoms, and Cr has almost no contribution to magnetism. The spin moment contributed by Fe is rather small, 0.6 $\mu_B$/Fe, and this small value quite well corroborates the weak itinerant character of the system. A small but non-zero orbital moment indicates that the spin orbit coupling associated with Fe is weak. The XMCD signal from the Cr-L-edge is two orders of magnitude lower than that of Fe-L-edge indicating that the magnetic moment primarily arises from the Fe-site. The small moment (= 0.006 $\mu_B$/Cr atom) observed at the Cr site could be induced by doped Fe atoms or can be induced by lattice defects.

\par
The critical analysis around  the FM Curie point of Cr$_{1.9}$Fe$_{0.1}$GeC preferably assigns a mean-field model for the phase transition. This magnetic MAX phase compound has a quasi-two-dimensional layered structure. We also observed indications of spin fluctuation. Such properties do not corroborate with a mean-field model of magnetic interaction. However, the magnetic interaction in this itinerant magnet is likely to be long-range in nature arising from the indirect  Ruderman–Kittel–Kasuya–Yosida (RKKY) type mechanism.  In case of a long-range magnetic interaction, the spin-spin correlation length is quite high. As a result, we can assume that a single spin can interact with all other spins of the system. This in turn produces a uniform lattice field on all spins, effectively reducing the system to obey a mean-field model~\cite{long_range}.
\par
In a previous work, Mn doping at the Cr site of Cr$_2$GeC was found to induce itinerant ferromagnetism~\cite{Liu2014}. Nuclear magnetic resonance~\cite{maniv} and powder neutron diffraction~\cite{rivin} indicate that the magnetism originates primarily from the doped Mn atoms. Compared to Mn, Fe doping gives rise to a slightly higher magnetic moment along with a higher ferromagnetic $T_C$. For example, Cr$_{1.9}$Mn$_{0.1}$GeC has a saturation moment of 0.068 $\mu_B$(f.u.)$^{-1}$ at 2 K and $T_C =$ 75 K, while Cr$_{1.9}$Fe$_{0.1}$GeC has $q_s =$ 0.083 $\mu_B$(f.u.)$^{-1}$ and $T_C =$ 246 K. The present work on Fe doping and previous studies on Mn substitution indicate that C$_2$GeC lies close to the paramagnetic / FM boundary. Although Cr is a 3$d$ transition metal, it does not induce any magnetic moment and long-range ordered state in Cr$_2$GeC. A small doping of post-Cr 3$d$ transition metal gives rise to itinerant ferromagnetism through the Stoner mechanism because of the enhanced Coulomb repulsion among band electrons~\cite{santiago}.   
\par
In conclusion, we have successfully induced ferromagnetism in the MAX phase compound Cr$_{2}$GeC by doping a few percent Fe at the Cr site. The resulting compositions show weak itinerant ferromagnetism and the critical exponents are found to obey the mean-field model, presumably arising from the long-range character of the magnetic interaction. Our microscopic study categorically identifies the doped Fe as the source of the ordered magnetic moment. The FM Curie temperatures of the Fe-doped samples are close to room temperature, and they can be useful magnetic MAX-phase materials for future spintronics applications. 
\section{Acknowledgment}
The authors acknowledge the European Synchrotron Radiation Facility (ESRF), France (proposal number HC5356), and  Hiroshima Synchrotron Radiation Center (HiSOR), Hiroshima University, Japan, for providing synchrotron radiation facilities.

\end{document}